\documentclass[aps,prl,reprint,groupedaddress,showpacs]{revtex4-1}

\usepackage{graphicx} 
\usepackage[colorlinks=true,citecolor=blue, linkcolor=blue, urlcolor=blue]{hyperref} 
\begin{document}

\title{Observation of Transient Momentum-Space Interference During Scattering of a Condensate From an Optical Barrier}

\author{Rockson Chang}
\email[]{rchang@physics.utoronto.ca}
\author{Shreyas Potnis}
\author{Christopher W. Ellenor}
\altaffiliation{Currently at the Department of Electrical Engineering, Stanford University, USA}
\author{Mirco Siercke}
\altaffiliation{Currently at the Centre for Quantum Technologies, Nanyang Technological University, Singapore}
\author{Alex Hayat}
\author{Aephraim M. Steinberg}
\affiliation{Department of Physics and Institute of Optics, University of Toronto, 60 St. George Street, Toronto, ON, M5S 1A7, Canada}
\date{\today}

\begin{abstract}
Scattering theory traditionally deals with the asymptotic behaviour of a system far removed from the actual scattering event.  Here we present an experimental study of the one-dimensional scattering of a non-interacting condensate of $^{87}$Rb atoms from a potential barrier in the non-asymptotic regime, for which the collision dynamics are still ongoing.  We show that for near-transparent barriers, there are two distinct transient scattering effects that arise and dramatically change the momentum distribution during the collision: a deceleration of wavepacket components in mid-collision, and an interference between incident and transmitted portions of the wavepacket.  Both effects lead to the re-distribution of momenta giving rise to a rich interference pattern that can be used to perform reconstruction of the single-particle phase profile.
\end{abstract}


\maketitle


Scattering is traditionally described asymptotically, connecting the initial and final scattering states, while bypassing the dynamics during the collision itself \cite{LL}.  In typical particle scattering experiments, measurements are made long after the collision and thus the asymptotic solutions provide an accurate description of the observations.  Yet, there are circumstances in which the scattering cannot be described asymptotically and knowledge of the full wavefunction is required, for example in understanding the quantum kinetics of moderately dense gases \cite{Snider90, Snider96}.  Brouard and Muga \cite{Muga98, Muga98AP} theoretically studied the one-dimensional scattering of a wavepacket from a delta-function potential.  They showed that during the collision with a repulsive potential, the momentum-space wavefunction could exhibit dramatically different features, for instance the generation of high-momentum components.  This non-classical momentum enhancement is a consequence of the wave nature of matter, in the spirit of quantum reflection \cite{Nayak83, Shimizu01, Pasquini04} - an effect that occurs when the potential changes abruptly on the scale of the de Broglie wavelength, irrespective of the sign of the change.  However, the effect described here is distinctly transient, manifesting itself only during the scattering event, and vanishing in the asymptotic limits.

Over the past few decades, impressive advances in experimental techniques have granted access to ever faster timescales and lower energies, allowing for direct probes of the non-asymptotic scattering regime.  For example, ultrafast laser pulses can now be used to probe sub-femtosecond timescales, providing time-resolved probes of electron dynamics \cite{Zewail00}.  In parallel, atom cooling techniques \cite{Lett88,VanDruten96, Ketterle99} have matured, and now routinely produce samples in the nano-Kelvin regime \cite{Cornell95, Ketterle95}.  The scattering dynamics of these ultracold systems occur on an easily accessible microsecond timescale, making them an ideal system for investigation of non-asymptotic scattering.  Bose-Einstein condensates of dilute gases have been used extensively to observe matter-wave phenomena \cite{Ketterle97dbleSlit, Phillips00, Pasquini04}.  Due to the relatively low densities, the inter-particle interactions are weak, allowing for clear observation of single-particle quantum effects.  Furthermore, the dynamic Stark effect allows for creation of nearly arbitrary potentials with spatial features limited only by the wavelength of light used to induce them \cite{Grimm99}, and can be modulated much faster than the dynamics of the condensate \cite{Boshier09}.  Here we make use of these advantages to experimentally study scattering of a matter wavepacket from an optically-induced potential barrier in the non-asymptotic regime.  We observe the momentum-space wavefunction reshaping effect predicted by Brouard and Muga, and further show that scattering from a finite-sized potential gives rise to a second effect due to the deceleration of wavepacket components in mid-collision.  These transient scattering effects are observed to result in a rich momentum-space interference pattern.

\begin{figure}
\centering
\includegraphics[width=0.65\columnwidth]{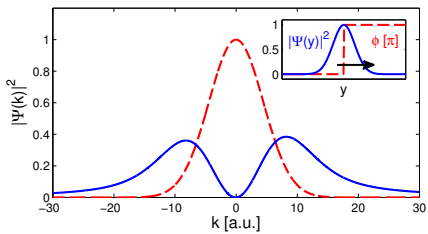}
\caption{\label{fig:Muga} Transient enhancement of momentum during scattering with a delta-function barrier.  Shown is the simulated momentum distribution, in the wavepacket centre-of-mass frame, long before/after the collision (red dashed line) and during the collision (blue solid line).  Inset: Position-space distribution during scattering (blue solid line), and phase-profile (red dashed line).  The barrier height has been chosen such that it results in a $\pi$-phase shift across the wavepacket.}
\end{figure}

\begin{figure}
\includegraphics[width=\columnwidth]{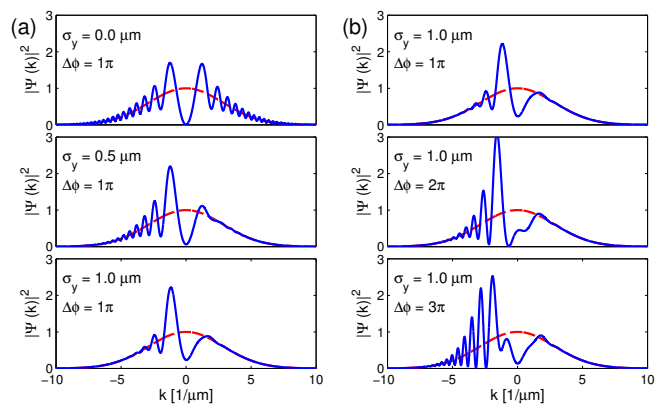}
\caption{\label{fig:realisticMuga} Momentum-space interference during scattering for a linearly chirped wavepacket.  Simulated momentum distributions, in the centre-of-mass frame of the incident wavepacket,  before/after the collision (red dashed line) and during the collision (blue solid line).  (a) The left column shows the dependence on barrier width.  For a delta-function barrier, T-I interference is dominant, creating a symmetric distribution.  For a finite-width barrier, P-I interference becomes dominant, breaking the symmetry.  (b) The right column shows the effect of increasing the barrier height.}
\end{figure}

At first glance, the predicted momentum redistribution and enhancement of high momenta might be interpreted as a spatial compression of the wavepacket as it impinges upon the barrier, resulting in a broadened momentum distribution.  Indeed, it has been shown that the enhancement increases with the height of the barrier, saturating in the opaque barrier limit \cite{Muga98AP}.  However Perez, Brouard and Muga \cite{Muga01} later showed that a much more significant momentum enhancement unexpectedly occurs for wavepackets with kinetic energy well above the barrier height.  They interpreted this effect as an interference between incident and transmitted portions of the wavepacket.  In the limit of a nearly transparent barrier, the primary effect on the wavepacket is to write a phase shift.  During the collision, half of the wavepacket has yet to reach the barrier and half has been transmitted.  Thus there exists a phase discontinuity dividing the incident and transmitted components of the wavepacket  in mid-collision (Fig.~\ref{fig:Muga} inset).  This sharp spatial feature in the phase results in destructive interference of the central momentum component and constructive interference in the tails \cite{Ruschhaupt09}.  The net result is a symmetrically broadened momentum distribution which exhibits non-classical momentum enhancement, and tails that fall off as $1/k^2$ (Fig.~\ref{fig:Muga}).  For an incident wavepacket with linear chirp, as would result from free expansion, the momentum distribution during scattering exhibits a much richer structure.  In mid-collision, the phase discontinuity is located at the spatial centre of the wavepacket.  Only the central momentum components, predominantly located near the centre of the spatial wavepacket, acquire the long momentum tails.  The high and low momentum components, which are spatially far from the discontinuity, remain unaffected.  The resulting momentum distribution displays rapid fringes due to interference between the redistributed central momentum component, and components in the tails of the distribution (Fig.~\hyperref[fig:realisticMuga]{\color{blue}\ref{fig:realisticMuga}(a)}). 

For finite-width barriers, the interference effect is significantly reduced.  This reduction depends exponentially on the product of the barrier width $\sigma_y$ and the wavepacket momentum width $\Delta k$.  In addition to the reduction of the momentum enhancement, there is a secondary effect which is that a centre-of-mass momentum is imparted.  Components of the wavepacket in mid-collision receive a net deceleration.  These components are effectively pushed to lower momenta, resulting in interference between the incident and pushed components of the wavepacket (Fig.~\ref{fig:realisticMuga}).  This breaks the symmetry of the momentum distribution and is distinct from the interference between the transmitted and incident components.  For our typical experimental parameters, $\sigma_y\cdot\Delta k\approx 4$, the transmitted-incident (T-I) interference effect is strongly suppressed, and the scattering is dominated by pushed-incident (P-I) interference.  However, since the latter only changes the distribution of slow-moving components of the wavepacket and the interference increases our sensitivity to small amplitudes, it is still possible to unambiguously observe the T-I interference effect on the fast-moving, transmitted components.  An additional consequence of a finite-width scattering potential is that the strict 2$\pi$-phase periodicity in T-I interference that would occur for a delta-function potential is washed out by the continuous range of phases written on the wavepacket.

It is important to emphasize the transient nature of both T-I and P-I interference effects.  For our implementation, the barrier height is typically 8 times smaller than the average kinetic energy of the wavepacket; there is no classical reflection and a probability of quantum reflection less than 10$^{-13}$.  Yet even in this near-transparent barrier limit, there is a dramatic modification of the momentum distribution \emph{during} interaction with the barrier, which afterwards vanishes, returning to the original distribution.  In contrast, the only classical effect would be a temporary decrease in the momentum of the fraction of atoms in mid-collision at any given moment.

We prepare a Bose-Einstein condensate of $^{87}$Rb atoms in the ground state of an optical dipole trap formed by the intersection of two focused 980 nm beams (red-detuned from the 780 nm D2 transition) oriented in the horizontal $x$-$z$ plane.  The trap is cylindrical, with approximate radial and axial trapping frequencies $\omega_x=\omega_y=2\pi\cdot300$ Hz and $\omega_z=2\pi\cdot100$ Hz, respectively.  In this trap, we prepare nearly pure condensates of about $10^5$ atoms.  After preparation, we abruptly turn off the optical trap, dropping the atoms under gravity ($y$-direction) onto an optically-induced barrier potential.  The barrier is positioned beneath our atoms such that the collision typically occurs after $t_{col}\sim$ 4 ms of free-fall, during which the wavepacket acquires a linear chirp.  The chirped wavepacket then collides with the barrier.  To study the momentum distribution during the collision, we abruptly turn the optical barrier off within 20 $\mu$s, freezing out the dynamics of the collision.  We then perform a long time-of-flight (TOF) expansion, followed by absorption imaging after a total time  $t_{tot}=$ 30 ms.  This TOF maps the momentum of the particles during the collision to their final position, such that the imaged position distribution is representative of the momentum distribution of the atoms during the collision.

The barrier potential is generated by a 780 nm beam blue-detuned from the $^{87}$Rb D2 transition by 150 GHz.  This beam is approximately gaussian in shape with a large aspect ratio, and propagates along the $x$-axis.  The rms barrier width along the direction of wavepacket propagation is $\sigma_y=$ 1.1 $\mu$m.  The beam has a size $\sigma_z\sim$ 10 $\mu$m transverse to the wavepacket propagation, and a Rayleigh range of $x_R\sim$ 20 $\mu$m.  $\sigma_z$ is comparable to the size of the expanding cloud at the time of collision.  We broaden the potential along $z$ by rapidly scanning the position of the barrier using an acouto-optic deflector (AOD), extending the length of the barrier to $\pm$60 $\mu$m.  This scan occurs at 100 kHz and is much faster than the motion of the atoms, resulting in a time-averaged potential \cite{Boshier09}.  We tailor the AOD scanning waveform to generate a time-averaged potential flat to within 1$\%$ of the barrier height, ensuring that the collision dynamics are essentially one-dimensional.  The barrier height is typically around $k_B \times1$ $\mu$K, where $k_B$ is Boltzmann's constant, and is chosen such that the effect of the barrier is to write a typical phase shift of approximately 3$\pi$, while remaining small compared to the typical wavepacket centre-of-mass kinetic energy of $k_B\times8$ $\mu$K acquired in free-fall.

Note that for an interacting system, such a phase imprinting would result in soliton formation \cite{Burger99, Burger02}.  The effect studied here is a single-particle interference effect.  The repulsive inter-particle interactions in the system are only relevant during the first millisecond of the experiment, driving the self-similar expansion of the cloud and thus determining the momentum distribution prior to collision.  The mean-field energy is rapidly converted to kinetic energy on a timescale of $1/\omega_y $ \cite{Castin96}, and is negligible during the collision itself.  This has been confirmed by comparing an interacting 3D Gross-Pitaevskii equation simulation to a non-interacting Schrodinger equation simulation of the collision with matched pre-collision momentum distributions.

\begin{figure} 
\centering
\includegraphics[width=1\columnwidth]{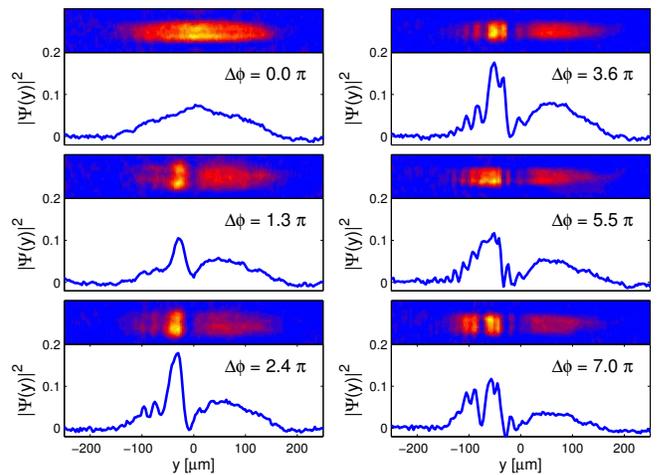}
\caption{\label{fig:data} Transient momentum-space interference during a collision.  Representative absorption images and corresponding 1D density profiles of a condensate after collision with a barrier at $t_{col}=$ 3.3 ms and TOF $t_{tot}=$ 30 ms.  The barrier height is expressed in terms of the phase shift imparted to fully transmitted atoms.  The cloud is moving under the influence of gravity in the $+y$-direction.  The images have been low-pass filtered in the transverse $z$-direction for presentation.}
\end{figure}

We measure the condensate momentum distribution during scattering for a variety of scattering parameters.  Figure~\ref{fig:data} shows a representative set of absorption images and the corresponding integrated 1D profiles for pre-collision expansion time $t_{col}=$ 3.3 ms and increasing barrier height.  The barrier height is expressed in terms of the estimated phase shift imparted to the fully transmitted atoms.  As the barrier height is increased we observe the development of a rich interference pattern, consistent with P-I interference.  For higher barriers, the central momentum component is more strongly decelerated, thus increasing the range over which interference is observed.  The fringe pattern that develops is reflective of the momentum-space phase profile of the condensate prior to collision.  For condensate expansion, we expect a quadratic phase profile \cite{Castin96}, and a fringe spacing that decreases linearly from cloud centre.  Note that our imaging system has a resolution of 3 $\mu$m, reducing the fringe visibility near the edges of the distribution.

\begin{figure}
\centering
\includegraphics[width=\columnwidth]{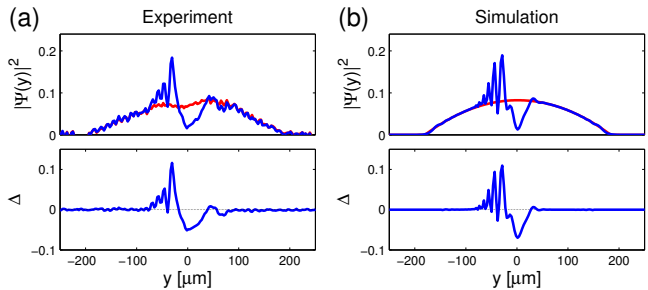}
\caption{\label{fig:TIdata} Evidence for transmitted-incident momentum-space interference for $t_{col}=$ 6.3 ms, $t_{tot}=$ 30 ms, and an estimated barrier phase shift of 3.1 $\pi$.  (a) Sample experimental 1D profile (average of 20 images), and (b) simulation (including image resolution effects).  Shown are the density profiles for un-collided (red) and collided clouds (blue), and the difference between these profiles, $\Delta$.  The pushed-incident interference is visible on the low-momentum side (negative $y$), while the transmitted-incident interference is visible on the high-momentum side (positive $y$) around $y=$ 50 $\mu$m.}
\end{figure}

Given the finite barrier width, T-I interference is strongly suppressed and the scattering is dominated by P-I interference.  However the latter effect is asymmetric, only affecting the low-momentum side of the distribution; thus T-I interference can still be observed on the high-momentum side.  In our parameter range, the effect is most visible for large phase curvatures, which is achieved with long pre-collision times.  Figure~\ref{fig:TIdata} shows a representative 1D profile (averaged over 20 images) for $t_{col}=$ 6.3 ms and an estimated barrier phase shift of 3.1$\pi$.  The experimental profiles show excellent agreement with our simulation, which accounts for the finite imaging resolution and camera pixel size.  Some jitter in the shot-to-shot position of the cloud may contribute to the reduced fringe visibility in the averaged experimental profiles when compared to simulation.  At this pre-collision time, we begin to see T-I interference fringes on the high-momentum side.  In principle, this type of interference also results in momentum enhancement; in our parameter regime, however, this effect is very small.  Though the T-I interference fringes are clearly observed, the expected momentum enhancement signal is too small to be quantitatively extracted from the experimental noise.  

\begin{figure}
\centering
\includegraphics[width=\columnwidth]{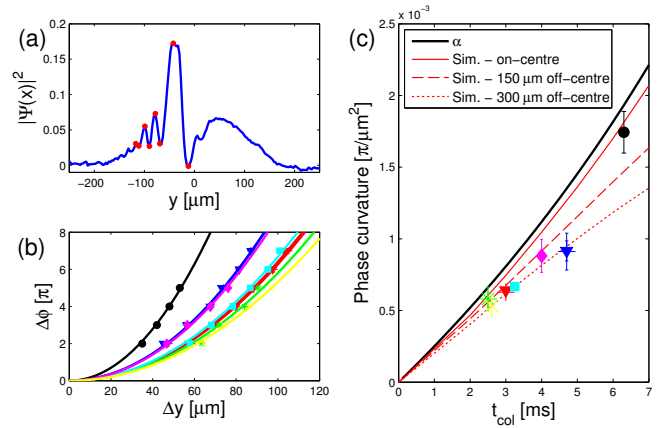}
\caption{\label{fig:phaseExtraction} Extracting the phase profile from experimental data.  (a) Sample profile for $t_{col}=$ 3.3 ms, where each peak/trough is marked and a relative $\pi$ phase is assigned.  (b) Each such phase profile is fitted to a quadratic.  (c) The fitted phase curvatures are plotted against pre-collision time.  The thick black line indicates expected free-expansion curvature, $\alpha$.  Red lines indicate the curvature extracted from simulated data for imaging on cloud centre, and increasingly off-centre (top to bottom; see text).  The colour and symbols of fringe data in (b) match that of the fitted curvatures in (c).}
\end{figure}

The observed momentum distributions during the collision exhibit a rich structure, containing both amplitude and phase information about the single-particle momentum-space wavefunction prior to collision.  Both T-I and P-I effects result in interference between momentum components initially at $k_i$ and scattered by the barrier at $t_{col}$ to $k_f$, and un-scattered components which have momentum $k_f$.  The interference depends on the relative phase accumulated along the respective trajectories.  By studying the fringe pattern we can extract the phase profile of the single-particle wavefunction.  To illustrate, we estimate the phase profile of our expanding condensate (Fig.~\ref{fig:phaseExtraction}) by marking the maxima and minima of each P-I interference feature in the observed profiles and assigning a relative $\pi$-phase shift between adjacent features.  For condensate expansion, a quadratic phase profile with a curvature that increases with the pre-collision time is expected for both momentum- and position-space distributions \cite{Castin96}.  The position-space phase profile after TOF is $\phi(z)=\alpha z^2$, where $\alpha\simeq\frac{m}{2\hbar}\left( \frac{1}{t_{tot}-t_{col}} - \frac{1}{t_{col}}\right)$.  We find close agreement between $\alpha$ and the phase curvature extracted from simulated data using the above technique.  

Our ability to focus on the atom cloud centre after TOF is limited to an accuracy of $\pm$300 $\mu$m.  Off-centre imaging results in distinct diffraction features, and a general reduction in extracted experimental fringe curvature from the expected value.  These imaging artifacts are well understood and entirely reproduced by our imaging simulations (performed after data collection).  Thus we compare our extracted fringe curvature to that predicted from our simulations for a range of imaging planes from on-centre to 300 um off-centre.  Note that the $t_{col}=$ 6.3 ms data set where the T-I interference effect is most visible (Fig.~\ref{fig:TIdata}), is consistent with imaging on cloud centre.

The technique used here to reconstruct the expanding condensate phase profile can be used for measurement of arbitrary single-particle phase profiles.  This interferometric approach bears some similarities to existing condensate phase profile reconstruction techniques \cite{Bigelow96, Phillips00}.  In contrast to these approaches, our technique in principle allows for single-shot reconstruction, provided the scattering potential is sharp enough.  For sharp scattering potentials, T-I interference dominates, symmetrically scattering the central momentum component across the cloud.  Thus the central momentum component acts like a local oscillator, beating against the other momentum components of the condensate simultaneously.  On the other hand, for broad scattering potentials the P-I interference dominates.  Assuming a symmetric phase profile, it is still possible to extract the full phase information, as we have shown for condensate expansion.  When using P-I interference, detailed knowledge of the scattering potential and its effect on the wavepacket is required to accurately recover the relative phase between momentum components.  Lastly, we note that since the fringe visibility is linked to the off-diagonal elements of the density matrix, this technique can be extended to perform full tomography on the single-particle state.  This could allow for the characterization of many body-correlations in ultracold systems, for example during the superfluid to Mott insulator transition.


The authors thank J. Gonzalo Muga for useful discussions; Ramon Ramos and Julian Schmidt for their assistance with the numerical simulations; and Ana Jofre and Marcelo Martinelli for their assistance in developing the Bose-Einstein condensation apparatus.  This work received financial support from CIFAR, NSERC, CFI, and ORDCF.  



%


\bibliography{Muga}
\end{document}